\documentclass[preprint,nofootinbib,floatfix,amsmath,amssymb,showpacs,superscriptaddress]{revtex4}

\usepackage{graphicx}
\usepackage{dcolumn}
\usepackage{bm}
\usepackage{color}

\begin{document}

\title{Vector and axial-vector couplings of D and D* mesons in 2+1 flavor Lattice QCD}

\author{K. U. Can}
\author{G. Erkol}
\affiliation{Laboratory for Fundamental Research, Ozyegin University, Nisantepe Mah. Orman Sok. No:13, Alemdag 34794 Cekmekoy, Istanbul Turkey}
\author{M. Oka}%
\affiliation{Department of Physics, H-27, Tokyo Institute of Technology, Meguro, Tokyo 152-8551 Japan}
\author{A. Ozpineci}%
\affiliation{Physics Department, Middle East Technical University, 06531 Ankara, Turkey}
\author{T. T. Takahashi}
\affiliation{Gunma National College of Technology, Maebashi, Gunma 371‐8530, Japan }

\date{\today}

\begin{abstract}
Using the axial-vector coupling and the electromagnetic form factors of the D and D$^\ast$ mesons in 2+1 flavor Lattice QCD, we compute the D$^\ast$D$\pi$, DD$\rho$ and D$^\ast$D$^\ast\rho$ coupling constants, which play an important role in describing the charm hadron interactions in terms of meson-exchange models. We also extract the charge radii of D and D$^\ast$ mesons and determine the contributions of the light and charm quarks separately. 

\end{abstract}
\maketitle

\section{Introduction}
An approach that utilizes the effective meson Lagrangian can successfully describe the suppression of $J/\psi$ production, which is considered to be a signal for the formation of quark-gluon plasma in relativistic heavy-ion collisions (RHIC)~\cite{Matinyan:1998cb, Haglin:1999xs, Lin:1999ad, Oh:2000qr}. This model has also been used to study $J/\psi$ absorption by nucleons~\cite{Liu:2001ce}, charm production from proton-proton collisions~\cite{Liu:2003be} and charm photoproduction off the nucleons~\cite{Liu:2003hi}. The effective Lagrangian employed in this method includes interaction vertices among $\pi$, $\rho$, $J/\psi$, D and D$^*$. The coupling constants of these mesons then play a crucial role in giving an accurate description of charm-hadron production and suppression in collisions performed at RHIC. 

In constructing the effective Lagrangian, meson exchange models use SU(4) symmetry together with Vector Meson Dominance model (VMD) to determine various coupling constants. While the DD$\pi$ coupling constant is small, the D$^\ast$D$^\ast\pi$ one is proportional to D$^\ast$D$\pi$ coupling constant according to heavy-quark spin symmetry~\cite{Oh:2000qr} and it is not necessarily small. A major contribution to processes involving diagonal transitions of D and D$^\ast$ mesons comes from $\rho$ exchange. Our primary aim here is to calculate the DD$\rho$ and D$^\ast$D$^\ast \rho$ coupling constants from first principles using 2+1 flavor Lattice QCD. As a byproduct, we obtain the electromagnetic form factor and the charge radii of D and D$^\ast$ mesons. In order to benchmark our simulations and compare with the available literature, we started our calculations with the well-known D$^\ast$D$\pi$ coupling constant, $g_{D^\ast D\pi}$. Therefore we give our results for $g_{D^\ast D\pi}$ also for illustrational purposes.

The calculation in this work is reminiscent of the pion (or kaon) electromagnetic form factor, which is considered to be a good observable to test QCD in a broad range of energy regime. There are also similarities between the $\rho$ (or $K^\ast$)~\cite{Hedditch:2007ex} and the D$^\ast$ electromagnetic form factors considered here. It has been found that the experimental data of the pion electromagnetic form factor at small momentum transfer are described quite successfully by the VMD \emph{ansatz}~\cite{Sakurai:69} 
\begin{equation}
	F_\pi(Q^2)=\frac{m_\rho^2}{m_\rho^2+Q^2}\frac{g_{\pi\pi\rho}}{g_\rho},
\end{equation}
where $m_\rho$ is the $\rho$-meson mass, $g_{\pi\pi\rho}$ and $g_\rho$ are the $\pi$-$\rho$ and $\rho$-photon coupling constants, respectively. While the $\rho$-meson is expected to dominate the electromagnetic current around the meson pole in the timelike region, VMD can describe the data quite accurately up to $Q^2\simeq$1~GeV$^2$ in the spacelike region~\cite{Bonnet:2004fr}. The monopole form,
\begin{equation}\label{monopole}
	F_\text{mon}(Q^2)=\frac{1}{1+Q^2/\Lambda^2}
\end{equation}
inspired by the VMD hypothesis (by assuming universality $g_{\pi\pi\rho}=g_\rho$), has also been used as a useful tool to fit the data to and predict the charge radius of the pion~\cite{Frezzotti:2008dr, Boyle:2008yd, Nguyen:2011ek}. It has been inferred from a compilation of experimental and theoretical results that the deviation from VMD starts above $Q^2\simeq$1.5~GeV$^2$~\cite{Huber:2008id}, which can be identified as the transition energy scale for pion from low-energy behavior to perturbative QCD. Since the D meson is much heavier than the pion, the transition is expected to occur at higher momentum transfers. Motivated by the success of VMD in describing the electromagnetic form factor of the pion, we use the same method to calculate the DD$\rho$ and D$^\ast$D$^\ast \rho$ coupling constants from lattice QCD data. 

D$^\ast$D$\pi$ coupling constant has been determined experimentally by CLEO Collaboration as $g_{D^\ast D\pi}=17.9 \pm 0.3 \pm 1.9$ and studied in the literature extensively. Therefore this observable can serve as a useful benchmark tool in this sector. While the results for $g_{D^\ast D\pi}$ from early QCD sum rules~\cite{Colangelo:1994es,Belyaev:1994zk,Colangelo:1997rp,Khodjamirian:1999hb} and potential model~\cite{Colangelo:1994jc} studies are well below the experimental value, those from lattice-QCD works are in good agreement with the experiment~\cite{Abada:2002xe,Becirevic:2009xp,Becirevic:2012pf}. Our calculations for $g_{D^\ast D\pi}$ here improve upon previous studies in several aspects, such as the lattice size and the number of sea-quark flavors.

Our work is organized as follows: In Section~\ref{sec2} we present the theoretical formalism of D and D$^\ast$ form factors together with the lattice techniques we have employed to extract them. In Section~\ref{sec3} we give and discuss our numerical results. Section~\ref{sec4} contains a summary of our findings.

\section{The formulation and the lattice simulations}\label{sec2}
We compute the meson matrix elements of the vector current $V_\mu=\frac{2}{3}\overline{c}\gamma_\mu c+\frac{2}{3}\overline{u}\gamma_\mu u-\frac{1}{3}\overline{d}\gamma_\mu d$, which can be written in the form
\begin{equation}\label{matel}
	\langle D(p^\prime)|V_\mu(q)| D(p)\rangle=(p+p^\prime)_\mu~[e_c F^c (Q^2) + e_q F^q (Q^2)]
\end{equation}
for the D meson. As for the spin-1 D$^*$ meson, we have
\begin{align}
	\begin{split}\label{matel2}
	\langle D^*(p^\prime,s^\prime)|V_\mu(q)| D^*(p,s)\rangle=&~\epsilon^{\prime\ast}_\tau(p^\prime,s^\prime)\left\{ \tilde{G}_1(Q^2)(p^\mu+p^{\mu\prime})g^{\tau\sigma}\right.\\
	&\left.+\tilde{G}_2(Q^2)(g^{\mu\sigma}q^\tau-g^{\mu\tau}q^\sigma)-\tilde{G}_3(Q^2)q^\tau q^\sigma\frac{(p^\mu+p^{\mu\prime})}{2m_{D^*}^2}\right\} \epsilon_\sigma(p,s),
	\end{split}
\end{align}
where $\epsilon$ and $\epsilon^\prime$ are the polarization vectors of the initial and final vector mesons, respectively. The form factors $\tilde{G}_{1,2,3}$ can be arranged in terms of Sachs electric, magnetic and quadrupole form factors as follows~\cite{Brodsky:1992px}:
\begin{align}
	\begin{split}
		&F_C(Q^2)=\tilde{G}_1(Q^2)+\frac{2}{3}\eta F_Q(Q^2)\\
		&F_M(Q^2)=\tilde{G}_2(Q^2)\\
		&F_Q(Q^2)=\tilde{G}_1(Q^2)-\tilde{G}_2(Q^2)+(1+\eta)\tilde{G}_3(Q^2)
	\end{split}
\end{align}		
where $\eta=Q^2/{4m_{D^*}^2}$.

Here we consider D$^+$ and D$^{\ast +}$ mesons therefore we take $e_q=1/3$ (anti\textendash $d$-quark) and $e_c=2/3$ ($c$-quark). We use the notation $q^2=-Q^2=(p-p^\prime)^2$ and, $F^c$ and $F^q$ are the vector form factor of D, where the external field couples to the $c$- and the $d$-quark in the D-meson respectively. In the limit of vanishing four-momentum transfer, we have $F^c (0)=F^q (0)=1$. Note that $F_\text{EM}(Q^2)=[e_c F^c (Q^2) + e_q F^q (Q^2)]$ gives the electromagnetic form factor of D and we have $F_\text{EM}(0)=1$ due to charge conservation. Similar constraints hold also for the D$^\ast$ electric form factor, $F^\ast_\text{EM}=F_C(Q^2)$.

The D$^*$D$\pi$ coupling constant, $g_{D^*D\pi}$, can be accessed via the transition matrix element $\langle D(p') | A^\mu(q) | D^*(p,s) \rangle$, where the axial-vector current for the light quark is given by $A_\mu = \bar{u} \gamma_5 \gamma_\mu d$. This matrix element can be parameterized with three form factors, $F_0(q^2)$, $F_1(q^2)$ and $F_2(q^2)$:  
\begin{align}
	\label{param} 
	\begin{split}
		\langle D(p') | A^\mu(q) | D^\ast(p,s) \rangle &= 2m_{V} F_0(q^2)\frac{\epsilon^s . q}{q^2} q^\mu \\
		&+ (m_D+m_{D^\ast}) F_1(q^2)[\epsilon^{s \mu} - \frac{\epsilon^s . q}{q^2} q^\mu] \\
		&+ F_2(q^2) \frac{\epsilon^s . q}{m_D+m_{D^\ast}}[p^\mu + p^{'\mu} - \frac{m_{D^\ast}^2-m_D^2}{q^2}q^\mu].
	\end{split}
\end{align}
PCAC relation and the VMD imply that the divergence of the axial-vector current $q_\mu A^\mu$ is dominated by a soft pion: 
\begin{equation}
	\label{vdm} \langle D(p') | q^\mu A^\mu(q) | D^\ast(p,s) \rangle = g_{D^\ast D\pi}\frac{\epsilon^s(p) . q}{m_\pi^2 - q^2} f_\pi m_\pi^2 + \ldots .
\end{equation}

We refer the reader to Ref.~\cite{Abada:2002xe} for the details of our calculations for $g_{D^\ast D\pi}$; here we just summarize the main steps. Rewriting the matrix element in terms of transferred and final momenta by defining $p^\mu = (p' + q)^\mu $, we can identify $g_{D^\ast D\pi}$ in terms of $F_1(0)$ and $F_2(0)$ as
\begin{align}
	\label{fcoup}
	g_{D^\ast D\pi}=\frac{1}{f_\pi}\left[(m_D+m_{D^\ast}) F_1(0) + (m_{D^\ast}-m_D) F_2(0)\right].
\end{align}
Defining 
\begin{equation}
	G_1(q^2) = \frac{m_{D^\ast}+m_D}{f_\pi}F_1(q^2) \quad , \quad G_2(q^2) = \frac{m_{D^\ast}-m_D}{f_\pi}F_2(q^2)
\end{equation}
and rearranging Eq. \eqref{fcoup} we write $g_{D^\ast D\pi}$ as 
\begin{equation}
	\label{rfcoup}
	g_{D^\ast D\pi}=G_1(0)\left(1 + \frac{G_2(0)}{G_1(0)}\right).
\end{equation} 

To extract the coupling constants, we compute the mesonic two-point,
\allowdisplaybreaks{
\begin{align}
	\begin{split}\label{twopcf}
	&\langle C(t; {\bf p})\rangle=\sum_{\bf x}e^{-i{\bf p}\cdot {\bf x}} \langle \text{vac} | T [\chi(x) \bar{\chi}(0)] | \text{vac}\rangle,\\
	&\langle C_{\mu\nu}(t; {\bf p})\rangle=\sum_{\bf x}e^{-i{\bf p}\cdot {\bf x}} \langle \text{vac} | T [\chi_\mu(x) \bar{\chi}_\nu(0)] | \text{vac}\rangle,
	\end{split}
\end{align}
}%
and three-point correlation functions,
\allowdisplaybreaks{
\begin{align}
	&\langle \tilde{C}^\alpha(t_2,t_1; {\bf p}^\prime, {\bf p})\rangle=-i\sum_{{\bf x_2},{\bf x_1}} e^{-i{\bf p}\cdot {\bf x_2}} e^{i{\bf q}\cdot {\bf x_1}} \langle \text{vac} | T [\chi(x_2) V^\alpha(x_1) \bar{\chi}(0)] | \text{vac}\rangle,\\
	&\langle \tilde{C}_{\mu\nu}^\alpha(t_2,t_1; {\bf p}^\prime, {\bf p})\rangle=-i\sum_{{\bf x_2},{\bf x_1}} e^{-i{\bf p}\cdot {\bf x_2}} e^{i{\bf q}\cdot {\bf x_1}} \langle \text{vac} | T [\chi_\mu(x_2) V^\alpha(x_1) \bar{\chi}_\nu(0)] | \text{vac}\rangle,\\
	&\langle \tilde{C}_{\mu\nu}(t_2,t_1; {\bf p}^\prime, {\bf p})\rangle=-i\sum_{{\bf x_2},{\bf x_1}} e^{-i{\bf p}\cdot {\bf x_2}} e^{i{\bf q}\cdot {\bf x_1}} \langle \text{vac} | T [\chi(x_2) A_\mu(x_1) \bar{\chi}_\nu(0)] | \text{vac}\rangle.
\end{align}
}%
The meson interpolating fields are given as
\begin{equation}
	\chi(x)=[\overline{d}(x)\gamma_5 c(x)], \qquad \chi_\mu(x)=[\overline{d}(x)\gamma_\mu c(x)].
\end{equation}
In our setup, the three momentum of the outgoing meson is automatically projected to zero momentum due to wall method, which is explained below; \emph{i.e.} ${\bf p}~^\prime={\bf 0}$.

In terms of the quark propagators $S(x,x^\prime)$, the three-point correlator for the D meson (we take the vector-field coupling as an example) can also be written as
\begin{align}\label{threepcf}
	\langle \tilde{C}^\alpha(t_2,t_1; {\bf p}^\prime, {\bf p})\rangle=-i\sum_{{\bf x_2},{\bf x_1}} e^{-i{\bf p}\cdot {\bf x_2}} e^{i{\bf q}\cdot {\bf x_1}} \\
	\times\langle \text{Tr}[\gamma_5~S_d(0,x_1)~\gamma^\alpha~S_d(x_1,x_2)~\gamma_5~S_c(x_2,0)]\rangle.
\end{align}
A similar expression holds also for the D$^*$ meson. While point-to-all propagators $S_d(0,x_1)$ and $S_c(x_2,0)$ can be easily obtained, the computation of all-to-all propagator $S_d(x_1,x_2)$ is a formidable task. One common method is to use a \emph{sequential source} composed of $S_d(0,x_1)$ and $S_c(x_2,0)$ for the Dirac matrix and invert it in order to compute $S_d(x_1,x_2)$~\cite{Wilcox:1991cq}. 
However, this method requires to fix sink operators before matrix inversions. 

An approach that does not require to fix sink operators in advance is the \emph{wall method}, where a summation over the spatial sites at the sink time point, ${\bf x_2}$, is made before the inversion. This corresponds to having a wall source or sink:
\begin{align}
	\begin{split}\label{threepcfw}
	\langle \tilde{C}^\alpha_{SW}(t_2,t_1; {\bf 0}, {\bf p})\rangle=-i\sum_{{\bf x_2}, {\bf x_2^\prime},{\bf x_1}} e^{i{\bf q}\cdot {\bf x_1}} \\
	\times\langle \text{Tr}[\gamma_5~S_d(0,x_1)~\gamma^\alpha~S_d(x_1,x_2^\prime)~\gamma_5~S_c(x_2,0)]\rangle
	\end{split}
\end{align}
where the propagator (instead of the hadron state) is projected on to definite momentum ($S$ and $W$ are smearing labels for \emph{shell} and \emph{wall}). Since the wall sink/source is a gauge-dependent object, one has to fix the gauge. Here we fix the gauge to Coulomb which produces a somewhat better coupling to the hadron ground state as compared to the Landau gauge. The wall method has the advantage that one can first compute the shell and wall propagators and then contract the propagators in order to obtain the three-point correlator, avoiding any sequential inversions. Use of the wall method allows us to compute the D and D$^\ast$ axial-vector and electromagnetic-transition channels simultaneously, which would require separate treatments with the traditional sequential-source method.

We compute the matrix element in Eq.~\eqref{matel} using the ratio
\begin{align}
	\begin{split}\label{ratio}
	&R^\alpha(t_2,t_1;{\bf p}^\prime,{\bf p};\mu)=\\
	&\quad\frac{\langle \tilde{C}^\alpha_{SW}(t_2,t_1; {\bf p}^\prime, {\bf p})\rangle}{\langle C_{SW}(t_2; {\bf p}^\prime)\rangle} \left[\frac{\langle C_{SS}(t_2-t_1; {\bf p})\rangle}{\langle C_{SS}(t_2-t_1; {\bf p}^\prime)\rangle} \frac{\langle C_{SS}(t_1; {\bf p}^\prime)\rangle \langle C_{SS}(t_2; {\bf p}^\prime)\rangle}{\langle C_{SS}(t_1; {\bf p})\rangle \langle C_{SS}(t_2; {\bf p})\rangle} \right]^{1/2}.
\end{split}
\end{align}
$t_1$ is the time when the vector field interacts with a quark and $t_2$ is the time when the final meson state is annihilated. The ratio in Eq.~(\ref{ratio}) reduces to the desired form when $t_2-t_1$ and $t_1\gg a$, {\it viz.}
\begin{equation}\label{desratio}
	R(t_2,t_1;{\bf 0},{\bf p};0)\xrightarrow[t_2-t_1\gg a]{t_1\gg a} \frac{(E_D+m_D)}{2\sqrt{E_D\,m_D}}\,[e_c F^c (Q^2) + e_q F^q (Q^2)],
\end{equation}
where $m_D$ and $E_D$ are the mass and the energy of the initial baryon. We apply a procedure of seeking plateau regions as a function of $t_1$ in the ratio \eqref{desratio} and calculating the vector form factors $F^c (Q^2)$ and $F^q (Q^2)$. We extract the D-meson mass from the two-point correlator with shell source and point sink, and use the dispersion relation to calculate the energy at each momentum transfer.

The matrix element in Eq.~\eqref{matel2} is computed using the ratio
\begin{align}
\begin{split}\label{ratio2}
	&R_{\mu\nu}^\alpha(t_2,t_1;{\bf p}^\prime,{\bf p};\mu)=\\
	&\quad\frac{\langle \tilde{C}^{\mu\alpha\nu}_{SW}(t_2,t_1; {\bf p}^\prime, {\bf p})\rangle}{\langle C^{\mu\nu}_{SW}(t_2; {\bf p}^\prime)\rangle} \left[\frac{\langle C^{\mu\nu}_{SS}(t_2-t_1; {\bf p})\rangle}{\langle C^{\mu\nu}_{SS}(t_2-t_1; {\bf p}^\prime)\rangle}\frac{\langle C^{\mu\nu}_{SS}(t_1; {\bf p}^\prime)\rangle \langle C^{\mu\nu}_{SS}(t_2; {\bf p}^\prime)\rangle}{\langle C^{\mu\nu}_{SS}(t_1; {\bf p})\rangle \langle C^{\mu\nu}_{SS}(t_2; {\bf p})\rangle} \right]^{1/2}.
\end{split}
\end{align}
This ratio gives
\begin{align}
	&R^0_{ii}=\frac{p_i^2}{3m_{D^*}\sqrt{E_{D^*}m_{D^*}}}F_Q(Q^2)+\frac{E_{D^*}+m_{D^*}}{2\sqrt{E_{D^*}m_{D^*}}}F_C(Q^2)\\
	&R^0_{jj}\Big{|}_{j\neq i}=-\frac{p_i^2}{6m_{D^*}\sqrt{E_{D^*}m_{D^*}}}F_Q(Q^2)+\frac{E_{D^*}+m_{D^*}}{2\sqrt{E_{D^*}m_{D^*}}}F_C(Q^2).
\end{align}
In order to single out the electric form factor we compute
\begin{equation}\label{ratsonDs}
	\frac{1}{3}\sum_{i=1,2,3} R_{ii}^0(t_2,t_1;{\bf 0}, p_j;0)\xrightarrow[t_2-t_1\gg a]{t_1\gg a} \frac{(E_{D^*}+m_{D^*})}{2\sqrt{E_{D^*}\,m_{D^*}}}\,[e_c F^{*c} (Q^2) + e_q F^{*q} (Q^2)].
\end{equation}

Finally, we compute the form factors $F_1(0)$ and $F_2(0)$ needed for $g_{D*D\pi}$ via the ratios~\cite{Abada:2002xe}
\begin{equation}
	\begin{aligned}
		\label{r1}
		R_1(t) =\frac{\tilde{C}_{SW}^{ii}(t) \sqrt{Z_{D^*}} \sqrt{Z_D}}{C_{SS}^{ii}(t) C_{ WW}(t_2-t_1)}, \qquad
		R_2(t) = \frac{\tilde{C}_{SW}^{10}(t, \vec{q}) \sqrt{Z_{D^*}} \sqrt{Z_D}}{C_{SS}^{22}(t_1,\vec{q}) C^{WW}(t_2-t_1)}, \\
		\\
		R_3(t) = \frac{\tilde{C}_{SW}^{11}(t, \vec{q}) \sqrt{Z_{D^*}} \sqrt{Z_D}}{C_{SS}^{22}(t_1,\vec{q}) C_{WW}(t_2-t_1)}, \qquad
		R_4(t) = \frac{\tilde{C}_{SW}^{22}(t, \vec{q}) \sqrt{Z_{D^*}} \sqrt{Z_D}}{C_{SS}^{22}(t_1,\vec{q}) C_{WW}(t_2-t_1)}.
	\end{aligned}	
\end{equation}
The masses and the normalization factors $Z_{D^*}$ and $Z_D$ are obtained from exponential fits to the zero-momentum two-point correlators, 
\begin{equation}
\label{eq:zvzp}
C(t_1;\vec{p}) \simeq Z_D\frac{e^{-E_D t_1}}{2 E_D}, \qquad C^{\mu\nu}(t_1;\vec{p}) \simeq Z_{D^*}\frac{e^{-E_{D^*} t_1}}{2 E_{D^*}}(\delta_{\mu \nu} - \frac{p^\mu p^\nu}{p^2}).
\end{equation}
$F_1(0)$ can be computed easily, however we should compute $F_2(0)$ by extrapolation since the term including $F_2(q^2)$ vanishes at zero momentum transfer:
\begin{align}
&	F_1(\vec{q} = 0,~t_1) = \frac{R_1(t_1)}{m_{D^*} + m_D}, \qquad F_1\left(\vec{q} = \frac{2 \pi}{L}( q_x, q_y, q_z),~t_1 \right) = \frac{R_4(t_1)}{m_{D^*}+m_D} \nonumber \\
	\nonumber \\
&	\frac{F_2}{F_1}(t_1) = \frac{(m_{D^*} + m_D)^2}{2 m_D^2 \vec{q}^{\;2}} \left[ \left(\vec{q}^{\;2} -E_{D^*}(E_{D^*} - m_D)\right) + \frac{m_{D^*} (E_{D^*} - m_D)}{E_{D^*}} \frac{R_3(t_1)}{R_4(t_1)} + i \frac{m_{D^*}^2 q_1}{E_{D^*}}\frac{R_2(t_1)}{R_4(t_1)} \right].
\end{align}
We assume the value of $F_2(0)$ to be close to its value at the smallest finite momentum transfer since we expect the $F_2 / F_1$ ratio to be insensitive to the changes of transferred momentum around the pion pole.

We make our simulations on a $32^3\times 64$ lattice with 2+1 flavors of dynamical quarks and the gauge configurations we use have been generated by the PACS-CS collaboration~\cite{Aoki:2008sm} with the nonperturbatively $O(a)$-improved Wilson quark action and the Iwasaki gauge action. We use the gauge configurations at $\beta=1.90$ with the clover coefficient $c_{SW}=1.715$, which give a lattice spacing of $a=0.0907(13)$ fm ($a^{-1}=2.176(31)$~GeV). The simulations are carried out with four different hopping parameters for the sea and the $u$,$d$ valence quarks, $\kappa_{sea},\kappa_{val}^{u,d}=$ 0.13700, 0.13727, 0.13754 and 0.13770, which correspond to pion masses of $\sim$ 700, 570, 410, and 300~MeV. The hopping parameter for the $s$ sea quark is fixed to $\kappa_{val}^{s}=0.1364$.

It is well known that the Clover action has discretization errors of $O(m_q\,a)$. Precision calculations such as the spectral properties and the hyperfine splittings may require removal or at least suppression of these lattice artefacts by considering improved actions such as Fermilab~\cite{ElKhadra:1996mp}. On the other hand, calculations which are insensitive to a change of charm-quark mass are less demanding in this respect~\cite{Bali:2011rd}. Considering also the precision levels we aim for the coupling constants and the fine spacing of our lattice, we choose to employ Clover action for the charm quark. We have checked the variation in our results by changing the charm-quark mass mildly and we confirmed that the coupling constants are insensitive to such a change (see the discussion below). Note that the Clover action we are employing here is a special case of the Fermilab heavy quark action with $c_{SW}=c_E=c_B$~\cite{Burch:2009az}. We determine the hopping parameter of the charm quark ($\kappa_{c}=0.1224$) so as to reproduce the mass of J/$\psi$. 

We employ smeared source and wall sink, which are separated by 12 lattice units in the temporal direction. Source operators are smeared in a gauge-invariant manner with the root mean square radius of $\sim 0.5$ fm. All the statistical errors are estimated via the jackknife analysis. We make our measurements on 45, 50, 50 and 70 configurations, respectively, for each quark mass. For the D-meson vector coupling, we make nine momentum insertions: $(|p_x|,|p_y|,|p_z|)=(0,0,0),(1,0,0),(1,1,0),(1,1,1),(2,0,0),(2,1,0),(2,1,1),(2,2,0),(2,2,1)$.  Given Eq.~\eqref{ratsonDs} for the D$^\ast$ meson, we are limited with the kinematics where the momentum should be inserted in one direction only. For the D$^\ast$ meson, we make our measurements with four momentum insertions: $(|p_x|,|p_y|,|p_z|)=(0,0,0),(1,0,0),(2,0,0),(3,0,0)$. We then rotate momentum in other directions and using the isotropy of space we average over equivalent momenta for both D and D$^\ast$ in order to increase the statistics. 

We consider point-split lattice vector current
\begin{equation}
V_\mu = 1/2[q(x+\mu)U^\dagger_\mu(1+\gamma_\mu)q(x) -q(x)U_\mu(1-\gamma_\mu)q(x+\mu)],
\end{equation}
which is conserved by Wilson fermions. Therefore it does not require any renormalization on the lattice. The local axial-vector current, on the other hand, needs to be renormalized on the lattice, where the renormalization factors are computed in a perturbative manner~\cite{AliKhan:2001tx}.

\section{Results and Discussion}\label{sec3}
We begin our discussion with the D$^\ast$D$\pi$ coupling constant. We find that the dominant contribution to the coupling constant comes from the first term in Eq.~\eqref{fcoup}. The second term including the ratio $G_2/G_1$ contributes to the coupling around $10\% $. Our results are listed in Table~\ref{tab:coup}. We give the dominant and minor contributions to the coupling constant individually.
\begin{table}[htb]
	\centering
	\begin{tabular}{cccc}
		\hline\hline
		$\kappa_{ud}$ & $G_1(q^2=0)$ & $G_2/G_1$ & $g_{D^*D\pi}$  \\
		\hline
		$0.13700$ & $14.15  (1.58)$ & $ 0.09(2) $ & $15.45 (1.78)$\\
		$0.13727$ & $13.63  (1.57)$ & $ 0.12(4)$ & $15.24 (1.81)$\\
		$0.13754$ & $12.76  (1.43)$ & $ 0.15(7)$ & $15.54 (2.08)$\\
		$0.13770$ & $15.46  (2.17)$ & $ 0.07(6)$ & $16.44 (2.41)$ \\
		\hline
		Lin. Fit &  &   &  $16.23(1.71)$   \\
		Quad. Fit &  &   & $17.09(3.23) $  \\		
		\hline\hline
	\end{tabular}
	\caption{Dominant and minor contributions to the coupling $g_{D^*D\pi}$ at each sea-quark mass we consider. }
	\label{tab:coup}
\end{table}

\begin{figure}[th]
	\centering
	\includegraphics[width=0.9\textwidth]{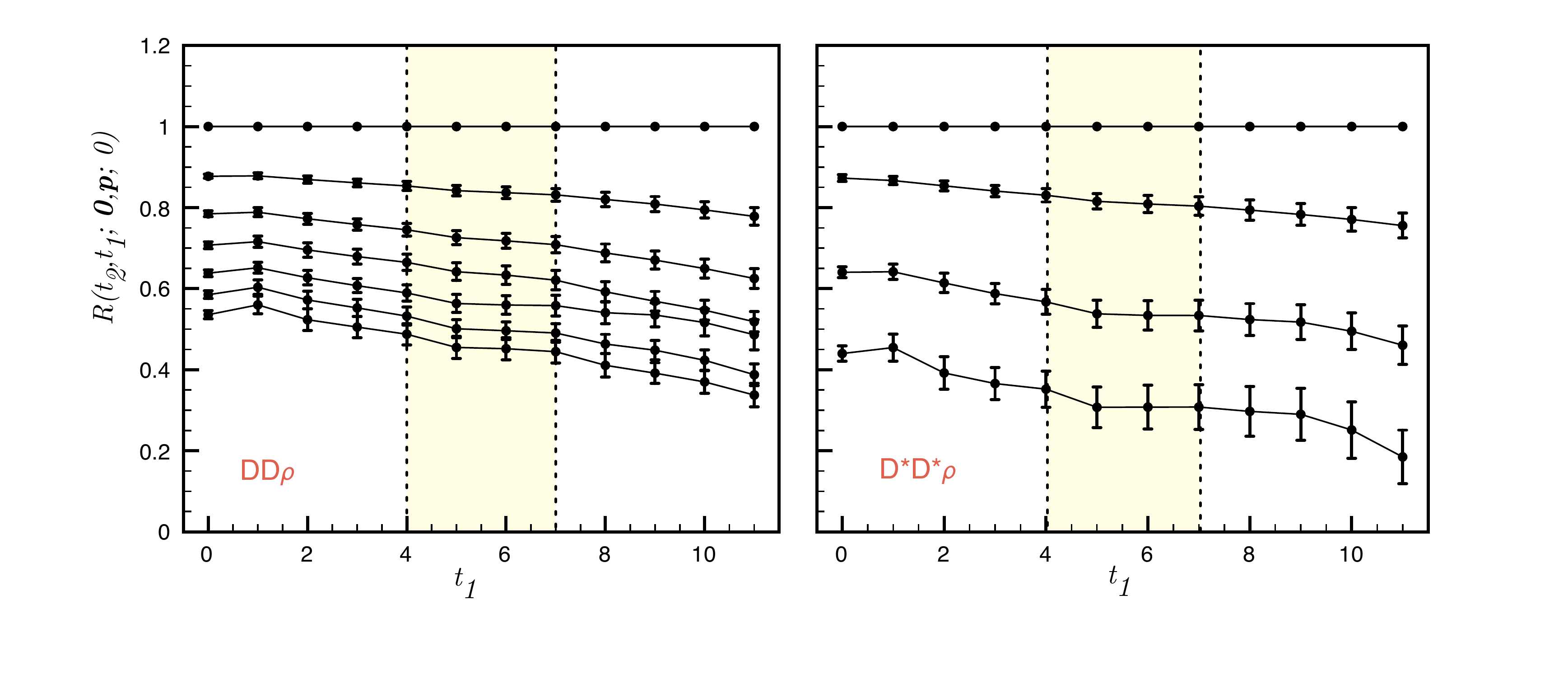}
	\caption{\label{plato} The ratio in Eq.~\eqref{ratio} (Eq.\eqref{ratio2}) as functions of the current insertion time, $t_1$, for $\kappa=0.13700$ and first seven (four) momentum insertions in the case of D (D$^\ast$).}
\end{figure}

As for the vector couplings, we construct the ratios in Eqs.~\eqref{ratio} and~\eqref{ratio2} in order to extract the form factors. To demonstrate the signal-to-noise ratio and show the chosen plateau regions, we give in Fig.~\ref{plato} the ratio in Eq.~\eqref{ratio} (Eq.\eqref{ratio2}) as functions of the current insertion time, $t_1$, for $\kappa=0.13700$ and for first seven (four) momentum insertions in the case of D (D$^\ast$). Plateaus appear around the middle of the source and the sink, $t_1\sim \frac{t_2}{2}$, which remain unchanged with larger source-sink separation.

To extract the coupling constant to the $\rho$-meson we use the VMD approach. There are different versions of VMD~(see Ref.~\cite{OConnell:1995wf} for a review). The two versions, which were first discussed by Sakurai~\cite{Sakurai:69}, differ by the mechanism the photon interacts with the hadron. In the more popular version, the photon is not allowed to directly couple to the hadron but only through a $\rho$-meson. This yields the following expression for the vector form factor:
\begin{equation}
	F_V(Q^2)=\frac{m_\rho^2}{m_\rho^2+Q^2}\frac{g_{DD\rho}}{g_\rho},
\end{equation}
where $g_\rho$ is a constant which determines the coupling of the vector meson to the photon. In this version in order to satisfy the constraint $F(0)=1$ one has to to assume $g_{DD\rho}=g_\rho$. In a second version of VMD, the photon can couple to both the hadron and the $\rho$ meson:
\begin{equation}\label{vmdform}
	F_V(Q^2)=\left[1-\frac{Q^2}{m_\rho^2+Q^2}\frac{g_{DD\rho}}{g_\rho}\right].
\end{equation}
In this version $F(0)=1$ is automatically satisfied and one does not need to assume $g_{DD\rho}=g_\rho$. We shall use the form in Eq.~\eqref{vmdform} to fit our data to and extract the coupling constants $g_{DD\rho}$ and $g_{D^*D^*\rho}$. We note that at the meson pole $Q^2\rightarrow -m_\rho^2$, both the VMD forms are in agreement giving $m_\rho^2/(Q^2+m_\rho^2)\frac{g_{DD\rho}}{g_\rho}$, so that the difference comes (if $g_{DD\rho}\neq g_\rho$) at $Q^2\rightarrow 0$ extrapolation. 

The coupling constant $g_\rho$ can be obtained from partial decay width of the $\rho$-meson to $e^+e^-$,
\begin{equation}
	\Gamma(V\rightarrow e^+ e^-)=\frac{4\pi}{3}\alpha^2 \frac{m_\rho}{g_\rho^2}
\end{equation}
with the fine structure constant $\alpha=1/137$. Using the experimental information~\cite{PhysRevD.86.010001} we find $g_\rho=4.96$. We shall neglect the contributions from the $\rho-\omega$ mixing as we have exact isospin symmetry on our lattice and such contributions are expected to play a role in the space-like region around the $\rho$-meson pole. 

We can obtain the electromagnetic charge radius of the D and D$^\ast$ mesons from the slope of the form factor at $Q^2=0$,
\begin{equation}
	\langle r^2 \rangle=-6 \frac{d}{dQ^2}F(Q^2)\bigg|_{Q^2=0}.
\end{equation}
For the monopole form in Eq.~\eqref{monopole} we have
\begin{equation}\label{emfitform}
	\langle r^2 \rangle=\frac{6}{\Lambda^2}.
\end{equation}
Inserting this expression back into Eq.~\eqref{monopole} for $\Lambda^2$ and rearranging we obtain
\begin{equation}\label{chradius}
	\langle r^2 \rangle=\frac{6}{Q^2}\left(\frac{1}{F(Q^2)}-1 \right).
\end{equation}
We extract the charge radii of the D and D$^\ast$ mesons using the above expression at the lowest finite momentum. This is a similar approach to that used in Ref.~\cite{Hedditch:2007ex} to calculate light meson electromagnetic form factors. We calculate the light-quark and charm-quark contributions to the charge radii separately. This information helps us to study the charge radii of individual quarks that are bound in the meson.

\begin{figure}[th]
	\centering
	\includegraphics[height=0.3\textheight]{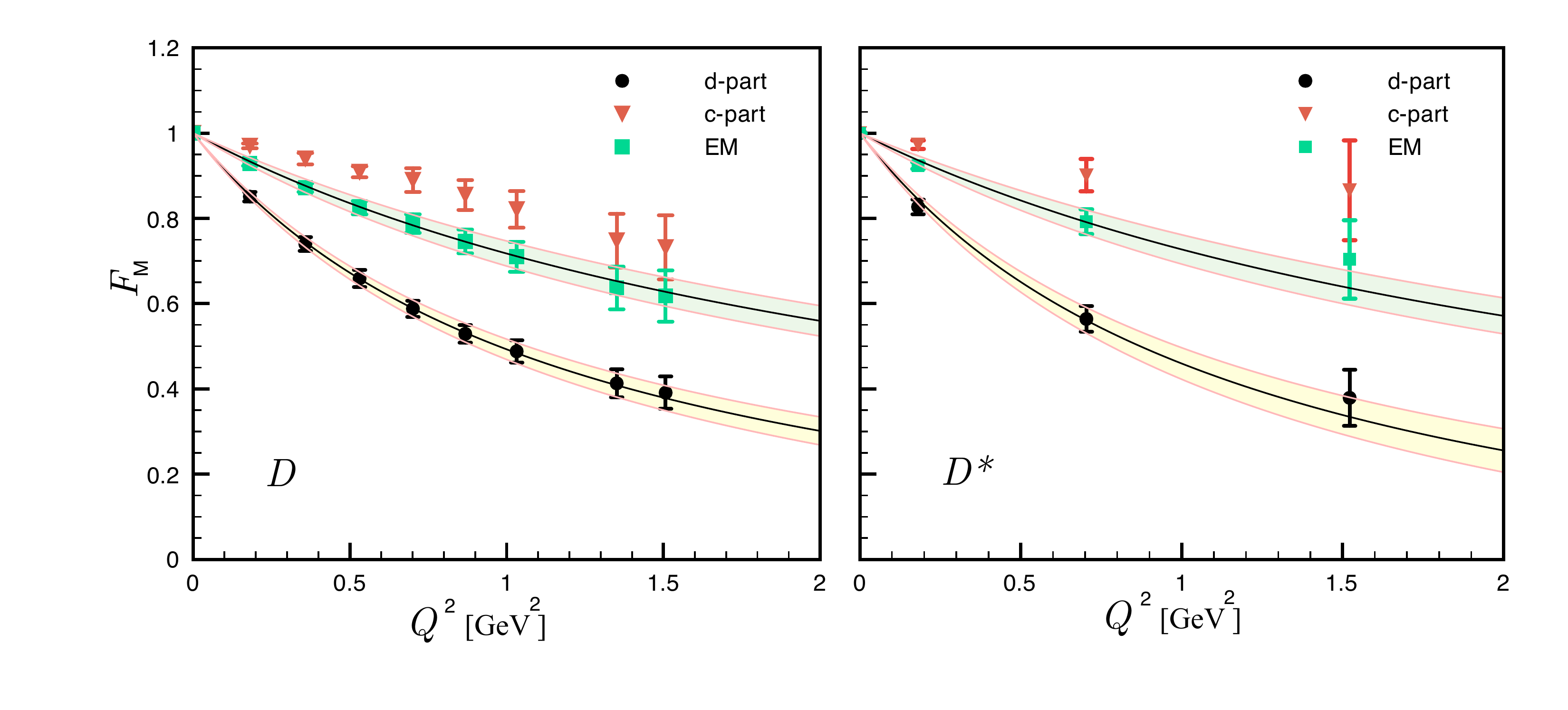}
	\caption{\label{ddrhoq2_cons} The form factors $F^{(\ast) d}(Q^2)$, $F^{(\ast) c}(Q^2)$ and the electromagnetic form factor $F_\text{EM}(Q^2)$ as a function of the four-momentum transfer $Q^2$ (in lattice units) for $\kappa=0.1370$. The shaded regions show jackknife error bars.}
\end{figure}	

In Fig.~\ref{ddrhoq2_cons} we show our lattice data for the form factors $F^{(*)d}(Q^2)$, $F^{(*)c}(Q^2)$ and the electromagnetic form factor $F_\text{EM}^{(*)}(Q^2)$ as functions of the four-momentum transfer $Q^2$ for $\kappa_{u,d}=0.1370$ with their jackknife errors. While we show data up to $Q^2\simeq 1.50$~GeV$^2$, we make our fits to first seven momentum-transfer values only~(up to $\simeq 1$~GeV$^2$). This is a region where VMD is expected to be valid. Nevertheless, the data at higher momentum transfers are also well described by the VMD form (for $F^{(*)d}(Q^2)$) and the monopole form (for the electromagnetic form factor) as can be seen in Fig.~\ref{ddrhoq2_cons}.

\begin{figure}[th]
	\includegraphics[width=0.7\textwidth]{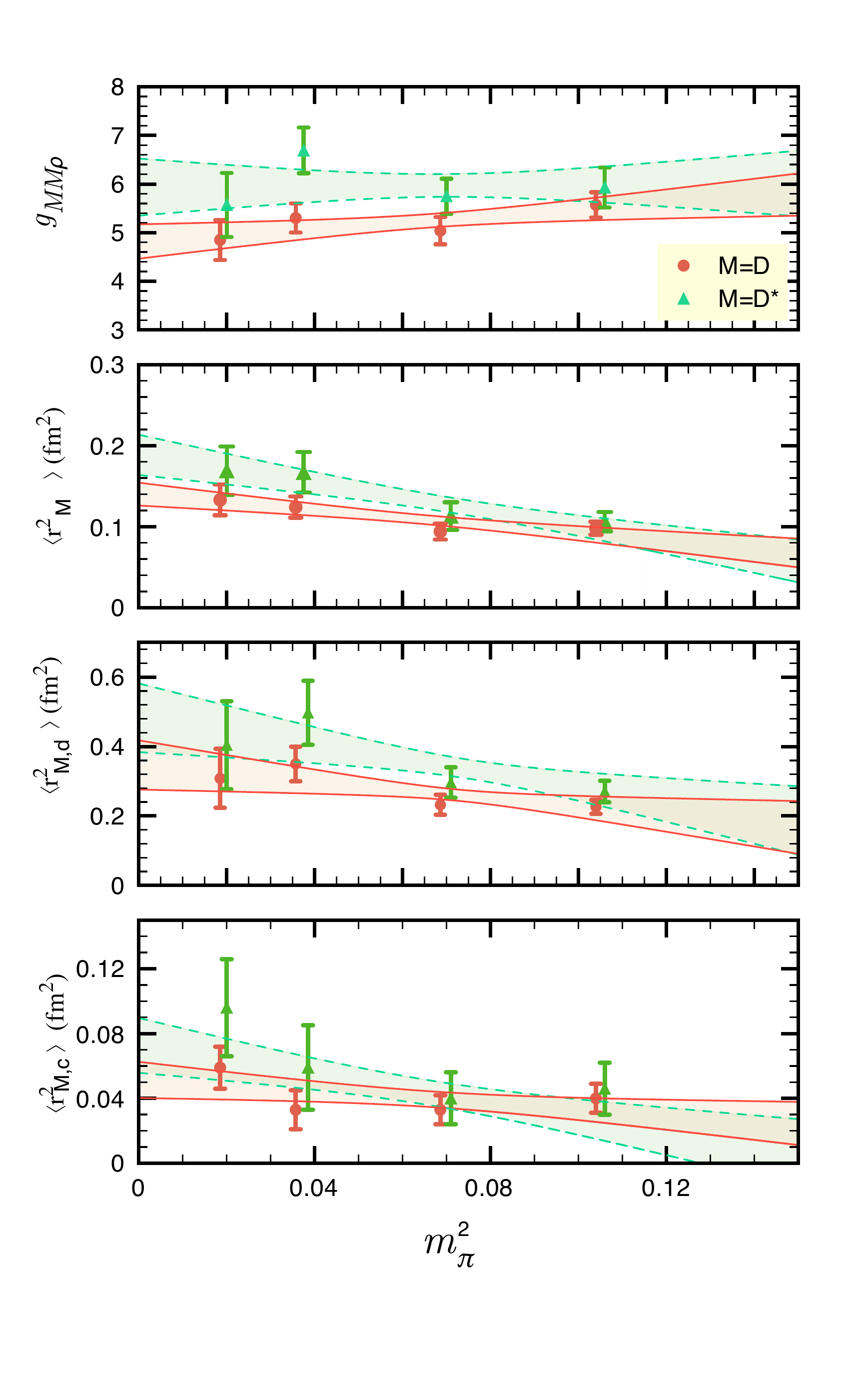}
	\caption{\label{chiral} $m_\pi^2$ dependence of the coupling constants $g_{DD\rho}$, $g_{D^\ast D^\ast\rho}$ and the charge radius $\langle r_{D^{^\ast}}^2\rangle$ together with quark-sector contributions. The shaded regions show the chiral extrapolations with a linear form and their jackknife errors.}
\end{figure}

Our numerical results are provided in Table~\ref{res_table}. We give the values of the coupling constants $g_{DD\rho}$ and $g_{D^\ast D^\ast\rho}$ at four different light quark masses and as determined by a fit to the VMD form in Eq.~\eqref{vmdform}. We also give the values of the charge radii of D and D$^\ast$ as obtained from Eq.~\eqref{chradius} together with contributions of light and charm quarks separately. We show the pion-mass dependence of the coupling constants and the charge radii in Fig.~\ref{chiral}.

In order to obtain the values of the observables at the chiral point, we perform linear and quadratic fits:
\begin{align}
	&f_\text{lin}=a\,m_\pi^2+b,\\
	&f_\text{quad}=a\,m_\pi^4+b\,m_\pi^2+c,
\end{align}
where $a,b,c$ are the fit parameters. For all the observables we calculate here, both the linear and quadratic fits give consistent results, whereas the linear fit produces smaller errors. Our chiral-extrapolated result for $g_{D^\ast D\pi}$, as can be seen in Table~\ref{tab:coup}, is in good agreement with experiment. Table~\ref{res_table} lists the chiral-extrapolated values of the meson coupling constants and charge radii. We observe that $g_{D^\ast D^\ast\rho}$  systematically lies above $g_{D D\rho}$ for all quark masses. According to heavy-quark spin symmetry, we have $g_{D D\rho}=g_{D^\ast D^\ast\rho}$ which provides a good test for the amount symmetry breaking. We obtain $(g_{D D\rho}/g_{D^\ast D^\ast\rho})_\text{lin}=0.814(53)$ and $(g_{D D\rho}/g_{D^\ast D^\ast\rho})_\text{quad}=0.888(91)$ at the chiral point, which indicate a breaking around 20\%. Comparing with $g_{D^\ast D\pi}$, SU(4) breaking is much larger while SU(4) symmetry predicts $g_{D D\rho}=g_{D^\ast D\pi}$. Our results for the coupling constants can be compared with those from QCD sum rules~\cite{Bracco:2001dj, Bracco:2007sg, Bracco:2011pg} as $g_{D D \rho }=2.9 \pm 0.4$ and $g_{D^\ast D^\ast \rho }=5.2 \pm 0.3$. Our value for $g_{D^\ast D^\ast\rho}$ is in agreement with that from QCD sum rules however there is a large discrepancy for $g_{D D\rho}$. On the other hand, our computed value for $g_{DD\rho}$ agrees well with that from Dyson-Schwinger equation studies in QCD as $g_{D D\rho}=5.05$~\cite{ElBennich:2011py}.

It is expected from the quark model that the hyperfine interaction between the quark and the antiquark is repulsive in the vector mesons and attractive for the pseudoscalar mesons. We observe from Table~\ref{res_table} that the charge radius of the D$^\ast$ meson is larger than that of D meson, indeed for all quark masses, in consistency with expectation from quark model. As it is seen in Fig.~\ref{chiral}, the coupling constants and the charge radii of D and D$^\ast$ mesons tend to be coincident as the quark mass increases. This is a natural result because the hyperfine interaction, which is of the form $\frac{\vec{\sigma}_Q\cdot\vec{\sigma}_q}{m_Q m_q}$, is reduced for a larger light-quark mass. We can also argue from this result that SU(4) breaking increases as the light-quark mass decreases. Indeed, similar relations hold for the $\rho$ coupling constants in the heavy quark limit as we have discussed above and one has $g_{DD\rho}=g_{D^\ast D^\ast \rho}$ (see Eq.(4.3) of Ref.~\cite{Casalbuoni:1992gi}).

Comparing the quark-sector contributions to the charge radii of D and D$^\ast$, we find that the dominant contribution comes from the light quark. This implies that the large mass of the $c$ quark suppresses the charge radii of D and D$^\ast$ mesons to smaller values compared to, \emph{e.g.}, charge radius of pion as $\langle r_\pi^2\rangle =$ 0.452~fm$^2$~\cite{PhysRevD.86.010001}. This is also in qualitative agreement with the conclusion of Ref.~\cite{Woloshyn:1985vd} that the meson size decreases as the quark mass increases and that heavier quarks have distributions of smaller radius. The available literature on the electromagnetic properties of the D and D$^\ast$ mesons is limited~\cite{ElBennich:2008qa,Hwang:2009qz}. Our result for the charge radius of the D meson is slightly below that from light-front quark model~\cite{Hwang:2009qz} as $\langle r^2_D \rangle=0.165^{-0.010}_{+0.011}$~fm$^2$.

\begin{table*}[ht]
	\caption{The coupling constants $g_{DD\rho}$, $g_{D^\ast D^\ast\rho}$, the charge radius of D and D$^\ast$ mesons together with individual quark-sector contributions and meson masses ($a\,m_\rho$ values are taken from~\cite{Aoki:2008sm}). We also give the D, D$^\ast$ and $J/\psi$ masses at different valence light quark masses. The chiral-extrapolated results are from linear and quadratic fits.  
}
\begin{center}
\begin{tabular*}{1.0\textwidth}{@{\extracolsep{\fill}}ccccc|cc}
		\hline\hline 
		$\kappa^{u,d}_{val}$   & $g_{DD\rho}$  &  $\langle r_{D}^2 \rangle$ (fm$^2$) & $\langle r_{D,d}^2 \rangle$ (fm$^2$) & $\langle r_{D,c}^2 \rangle$ (fm$^2$) & $a \,m_D$ & $a\,m_{J/\psi}$ \\
		\hline \hline
		0.13700 & 5.57(27) &  0.098(8) & 0.226(20) & 0.040(9) & 0.944(5) &1.453(5)\\
		0.13727 & 5.03(28) &  0.094(10) & 0.232(29) & 0.033(9) & 0.919(4) &1.447(3)\\
		0.13754 & 5.30(30) &  0.124(13) & 0.350(50) & 0.033(12) & 0.901(6) &1.434(6)\\
		0.13770 & 4.85(41) &  0.133(19) & 0.308(85)& 0.059(13) & 0.896(10) & 1.425(5) \\
		\hline
		Lin. Fit & 4.84(34) &  0.138(13) & 0.342(67)& 0.051(11) &  &  \\
		Quad. Fit & 4.90(56) &  0.152(26) & 0.320(118)& 0.074(16) &  &  \\
		\hline\hline 
		$\kappa^{u,d}_{val}$   & $g_{D^*D^*\rho}$  &  $\langle r_{D^\ast}^2 \rangle$ (fm$^2$) & $\langle r_{D^\ast,d}^2 \rangle$ (fm$^2$) & $\langle r_{D^*,c}^2 \rangle$ (fm$^2$) & $a\,m_{D^*}$ & $a\,m_\rho$ \\
		\hline \hline
		0.13700 & 5.93(41)  & 0.106(12) & 0.270(31) &  0.035(13) &  1.006(7)  & 0.5060(30)\\
		0.13727 & 5.75(36)  & 0.113(17) & 0.296(44) & 0.036(14) & 0.981(8) & 0.4566(36)\\
		0.13754 & 6.69(47)  & 0.167(25) & 0.497(92) & 0.044(21) & 0.971(7) & 0.4108(31)\\
		0.13770 & 5.57(66)  & 0.169(30) & 0.404(127) & 0.075(26) & 0.940(9) & 0.3895(94)\\
		\hline
		Lin. Fit & 5.94(56) &  0.185(24) & 0.475(94)& 0.071(16) &  &  \\
		Quad. Fit & 5.42(94) &  0.192(43) & 0.406(156)& 0.096(29) &  &  \\
		\hline\hline
\end{tabular*}
	\label{res_table}
\end{center}
\end{table*}

A few comments on the systematic errors are in order now. To check the validity of the use of Clover action for the charm-quark in the case of coupling constants, we have repeated our measurements for $\kappa_c=0.1216$ and $\kappa_c=0.1232$, which correspond to a change of $\sim\pm 100$~MeV in the charmonium mass. We have found that this leads to a change of less than 2\% in the coupling constants as well as in the charge radii. Then our results are practically insensitive to a mild change in the charm-quark mass justifying the validity of Clover action in this case.

For the finite-volume effect, the present spatial lattice extent is 32 units and the pion mass ranges from 0.136 to 0.322 in lattice unit, which gives $4.3\leq m_\pi L \leq 10.3$. A rule of thumb is that serious  effects seem to appear only when $m_\pi L \leq 4$; therefore we expect small finite-volume effects in our present calculations. 

We calculate only connected diagrams in this work. This means we neglect the effect of the disconnected diagram, where the external field is inserted in a sea-quark loop which in turn interacts with the meson two-point diagram with a gluon. The contribution of the disconnected diagrams is expected to cancel as long as we have an isovector field coupled to the meson as in the case of rho or pion couplings. A direct calculation of such diagrams appearing here only in the case of electromagnetic form factor, which has an isoscalar part as well as an isovector part, is difficult and numerically demanding. Their contribution has been found to be consistent with zero in the case of nucleon electric form factors~\cite{Alexandrou:2012zz}. We expect them to give negligible contributions also in the case charm-meson electromagnetic form factors. We also note that at higher-momentum transfers the effect of disconnected diagrams is further suppressed as a result of weaker coupling constant. 

\section{Conclusion}\label{sec4}

Using the axial-vector coupling and the electromagnetic form factor of the D and D$^\ast$ mesons, we have computed the D$^\ast$D$\pi$, DD$\rho$ and D$^\ast$D$^\ast\rho$ coupling constants, which play an important role in describing the charm-hadron interactions in terms of meson-exchange models. We have also extracted the charge radii of D and D$^\ast$ mesons and determined the contributions of light and charm quarks separately. Our final results for the coupling constants as linearly extrapolated to the chiral point are
\begin{equation}
	g_{D^\ast D \pi}=16.23(1.71),\qquad
	g_{D D \rho}=4.84(34),\qquad
	g_{D^\ast D^\ast \rho}=5.94(56).
\end{equation}
We have discussed SU(4) and heavy-quark spin symmetry breaking. We have found that the SU(4) breaking gets larger as we decrease the light-quark mass. We have also calculated the charge radii of D and D$^\ast$ mesons and found that the dominant contribution to the charge radius comes from the light quark. We have found that the large mass of the c quark suppresses the charge radii of D and D$^\ast$ mesons to smaller values as compared to the charge radius of pion.

\acknowledgments
We thank F. S. Navarra for discussions on QCD sum rules results in Ref.\cite{Bracco:2011pg}. All the numerical calculations in this work were performed on National Center for High Performance Computing of Turkey (Istanbul Technical University) under project number 10462009. The unquenched gauge configurations employed in our analysis were generated by PACS-CS collaboration~\cite{Aoki:2008sm}. We used a modified version of Chroma software system~\cite{Edwards:2004sx}. This work is supported in part by The Scientiﬁc and Technological Research Council of Turkey (TUBITAK) under project number 110T245 and in part by KAKENHI under Contract Nos. 22105503, 24540294 and 22105508.


\begin{thebibliography}{41}
\expandafter\ifx\csname natexlab\endcsname\relax\def\natexlab#1{#1}\fi
\expandafter\ifx\csname bibnamefont\endcsname\relax
  \def\bibnamefont#1{#1}\fi
\expandafter\ifx\csname bibfnamefont\endcsname\relax
  \def\bibfnamefont#1{#1}\fi
\expandafter\ifx\csname citenamefont\endcsname\relax
  \def\citenamefont#1{#1}\fi
\expandafter\ifx\csname url\endcsname\relax
  \def\url#1{\texttt{#1}}\fi
\expandafter\ifx\csname urlprefix\endcsname\relax\def\urlprefix{URL }\fi
\providecommand{\bibinfo}[2]{#2}
\providecommand{\eprint}[2][]{\url{#2}}

\bibitem[{\citenamefont{Matinyan and Muller}(1998)}]{Matinyan:1998cb}
\bibinfo{author}{\bibfnamefont{S.~G.} \bibnamefont{Matinyan}} \bibnamefont{and}
  \bibinfo{author}{\bibfnamefont{B.}~\bibnamefont{Muller}},
  \bibinfo{journal}{Phys.Rev.} \textbf{\bibinfo{volume}{C58}},
  \bibinfo{pages}{2994} (\bibinfo{year}{1998}), \eprint{nucl-th/9806027}.

\bibitem[{\citenamefont{Haglin}(2000)}]{Haglin:1999xs}
\bibinfo{author}{\bibfnamefont{K.~L.} \bibnamefont{Haglin}},
  \bibinfo{journal}{Phys.Rev.} \textbf{\bibinfo{volume}{C61}},
  \bibinfo{pages}{031902} (\bibinfo{year}{2000}), \eprint{nucl-th/9907034}.

\bibitem[{\citenamefont{Lin and Ko}(2000)}]{Lin:1999ad}
\bibinfo{author}{\bibfnamefont{Z.-w.} \bibnamefont{Lin}} \bibnamefont{and}
  \bibinfo{author}{\bibfnamefont{C.}~\bibnamefont{Ko}},
  \bibinfo{journal}{Phys.Rev.} \textbf{\bibinfo{volume}{C62}},
  \bibinfo{pages}{034903} (\bibinfo{year}{2000}), \eprint{nucl-th/9912046}.

\bibitem[{\citenamefont{Oh et~al.}(2001)\citenamefont{Oh, Song, and
  Lee}}]{Oh:2000qr}
\bibinfo{author}{\bibfnamefont{Y.-s.} \bibnamefont{Oh}},
  \bibinfo{author}{\bibfnamefont{T.}~\bibnamefont{Song}}, \bibnamefont{and}
  \bibinfo{author}{\bibfnamefont{S.~H.} \bibnamefont{Lee}},
  \bibinfo{journal}{Phys.Rev.} \textbf{\bibinfo{volume}{C63}},
  \bibinfo{pages}{034901} (\bibinfo{year}{2001}), \eprint{nucl-th/0010064}.

\bibitem[{\citenamefont{Liu et~al.}(2002)\citenamefont{Liu, Ko, and
  Lin}}]{Liu:2001ce}
\bibinfo{author}{\bibfnamefont{W.}~\bibnamefont{Liu}},
  \bibinfo{author}{\bibfnamefont{C.~M.} \bibnamefont{Ko}}, \bibnamefont{and}
  \bibinfo{author}{\bibfnamefont{Z.~W.} \bibnamefont{Lin}},
  \bibinfo{journal}{Phys.Rev.} \textbf{\bibinfo{volume}{C65}},
  \bibinfo{pages}{015203} (\bibinfo{year}{2002}).

\bibitem[{\citenamefont{Liu et~al.}(2003{\natexlab{a}})\citenamefont{Liu, Ko,
  and Lee}}]{Liu:2003be}
\bibinfo{author}{\bibfnamefont{W.}~\bibnamefont{Liu}},
  \bibinfo{author}{\bibfnamefont{C.~M.} \bibnamefont{Ko}}, \bibnamefont{and}
  \bibinfo{author}{\bibfnamefont{S.~H.} \bibnamefont{Lee}},
  \bibinfo{journal}{Nucl.Phys.} \textbf{\bibinfo{volume}{A728}},
  \bibinfo{pages}{457} (\bibinfo{year}{2003}{\natexlab{a}}),
  \eprint{nucl-th/0308013}.

\bibitem[{\citenamefont{Liu et~al.}(2003{\natexlab{b}})\citenamefont{Liu, Lee,
  and Ko}}]{Liu:2003hi}
\bibinfo{author}{\bibfnamefont{W.}~\bibnamefont{Liu}},
  \bibinfo{author}{\bibfnamefont{S.~H.} \bibnamefont{Lee}}, \bibnamefont{and}
  \bibinfo{author}{\bibfnamefont{C.-M.} \bibnamefont{Ko}},
  \bibinfo{journal}{Nucl.Phys.} \textbf{\bibinfo{volume}{A724}},
  \bibinfo{pages}{375} (\bibinfo{year}{2003}{\natexlab{b}}),
  \eprint{nucl-th/0302024}.

\bibitem[{\citenamefont{Hedditch et~al.}(2007)\citenamefont{Hedditch, Kamleh,
  Lasscock, Leinweber, Williams et~al.}}]{Hedditch:2007ex}
\bibinfo{author}{\bibfnamefont{J.}~\bibnamefont{Hedditch}},
  \bibinfo{author}{\bibfnamefont{W.}~\bibnamefont{Kamleh}},
  \bibinfo{author}{\bibfnamefont{B.}~\bibnamefont{Lasscock}},
  \bibinfo{author}{\bibfnamefont{D.}~\bibnamefont{Leinweber}},
  \bibinfo{author}{\bibfnamefont{A.}~\bibnamefont{Williams}},
  \bibnamefont{et~al.}, \bibinfo{journal}{Phys.Rev.}
  \textbf{\bibinfo{volume}{D75}}, \bibinfo{pages}{094504}
  (\bibinfo{year}{2007}), \eprint{hep-lat/0703014}.

\bibitem[{\citenamefont{Sakurai}(1969)}]{Sakurai:69}
\bibinfo{author}{\bibfnamefont{J.~J.} \bibnamefont{Sakurai}},
  \emph{\bibinfo{title}{Currents and mesons}} (\bibinfo{publisher}{University
  of Chicago Press}, \bibinfo{address}{Chicago}, \bibinfo{year}{1969}).

\bibitem[{\citenamefont{Bonnet et~al.}(2005)\citenamefont{Bonnet, Edwards,
  Fleming, Lewis, and Richards}}]{Bonnet:2004fr}
\bibinfo{author}{\bibfnamefont{F.~D.} \bibnamefont{Bonnet}},
  \bibinfo{author}{\bibfnamefont{R.~G.} \bibnamefont{Edwards}},
  \bibinfo{author}{\bibfnamefont{G.~T.} \bibnamefont{Fleming}},
  \bibinfo{author}{\bibfnamefont{R.}~\bibnamefont{Lewis}}, \bibnamefont{and}
  \bibinfo{author}{\bibfnamefont{D.~G.} \bibnamefont{Richards}}
  (\bibinfo{collaboration}{Lattice Hadron Physics Collaboration}),
  \bibinfo{journal}{Phys.Rev.} \textbf{\bibinfo{volume}{D72}},
  \bibinfo{pages}{054506} (\bibinfo{year}{2005}), \eprint{hep-lat/0411028}.

\bibitem[{\citenamefont{Frezzotti et~al.}(2009)\citenamefont{Frezzotti, Lubicz,
  and Simula}}]{Frezzotti:2008dr}
\bibinfo{author}{\bibfnamefont{R.}~\bibnamefont{Frezzotti}},
  \bibinfo{author}{\bibfnamefont{V.}~\bibnamefont{Lubicz}}, \bibnamefont{and}
  \bibinfo{author}{\bibfnamefont{S.}~\bibnamefont{Simula}}
  (\bibinfo{collaboration}{ETM Collaboration}), \bibinfo{journal}{Phys.Rev.}
  \textbf{\bibinfo{volume}{D79}}, \bibinfo{pages}{074506}
  (\bibinfo{year}{2009}), \eprint{0812.4042}.

\bibitem[{\citenamefont{Boyle et~al.}(2008)\citenamefont{Boyle, Flynn, Juttner,
  Kelly, de~Lima et~al.}}]{Boyle:2008yd}
\bibinfo{author}{\bibfnamefont{P.}~\bibnamefont{Boyle}},
  \bibinfo{author}{\bibfnamefont{J.}~\bibnamefont{Flynn}},
  \bibinfo{author}{\bibfnamefont{A.}~\bibnamefont{Juttner}},
  \bibinfo{author}{\bibfnamefont{C.}~\bibnamefont{Kelly}},
  \bibinfo{author}{\bibfnamefont{H.~P.} \bibnamefont{de~Lima}},
  \bibnamefont{et~al.}, \bibinfo{journal}{JHEP}
  \textbf{\bibinfo{volume}{0807}}, \bibinfo{pages}{112} (\bibinfo{year}{2008}),
  \eprint{0804.3971}.

\bibitem[{\citenamefont{Nguyen et~al.}(2011)\citenamefont{Nguyen, Ishikawa,
  Ukawa, and Ukita}}]{Nguyen:2011ek}
\bibinfo{author}{\bibfnamefont{O.~H.} \bibnamefont{Nguyen}},
  \bibinfo{author}{\bibfnamefont{K.-I.} \bibnamefont{Ishikawa}},
  \bibinfo{author}{\bibfnamefont{A.}~\bibnamefont{Ukawa}}, \bibnamefont{and}
  \bibinfo{author}{\bibfnamefont{N.}~\bibnamefont{Ukita}},
  \bibinfo{journal}{JHEP} \textbf{\bibinfo{volume}{1104}}, \bibinfo{pages}{122}
  (\bibinfo{year}{2011}), \eprint{1102.3652}.

\bibitem[{\citenamefont{Huber et~al.}(2008)}]{Huber:2008id}
\bibinfo{author}{\bibfnamefont{G.}~\bibnamefont{Huber}} \bibnamefont{et~al.}
  (\bibinfo{collaboration}{Jefferson Lab}), \bibinfo{journal}{Phys.Rev.}
  \textbf{\bibinfo{volume}{C78}}, \bibinfo{pages}{045203}
  (\bibinfo{year}{2008}), \eprint{0809.3052}.

\bibitem[{\citenamefont{Colangelo
  et~al.}(1994{\natexlab{a}})\citenamefont{Colangelo, Nardulli, Deandrea,
  Di~Bartolomeo, Gatto et~al.}}]{Colangelo:1994es}
\bibinfo{author}{\bibfnamefont{P.}~\bibnamefont{Colangelo}},
  \bibinfo{author}{\bibfnamefont{G.}~\bibnamefont{Nardulli}},
  \bibinfo{author}{\bibfnamefont{A.}~\bibnamefont{Deandrea}},
  \bibinfo{author}{\bibfnamefont{N.}~\bibnamefont{Di~Bartolomeo}},
  \bibinfo{author}{\bibfnamefont{R.}~\bibnamefont{Gatto}},
  \bibnamefont{et~al.}, \bibinfo{journal}{Phys.Lett.}
  \textbf{\bibinfo{volume}{B339}}, \bibinfo{pages}{151}
  (\bibinfo{year}{1994}{\natexlab{a}}), \eprint{hep-ph/9406295}.

\bibitem[{\citenamefont{Belyaev et~al.}(1995)\citenamefont{Belyaev, Braun,
  Khodjamirian, and Ruckl}}]{Belyaev:1994zk}
\bibinfo{author}{\bibfnamefont{V.}~\bibnamefont{Belyaev}},
  \bibinfo{author}{\bibfnamefont{V.~M.} \bibnamefont{Braun}},
  \bibinfo{author}{\bibfnamefont{A.}~\bibnamefont{Khodjamirian}},
  \bibnamefont{and} \bibinfo{author}{\bibfnamefont{R.}~\bibnamefont{Ruckl}},
  \bibinfo{journal}{Phys.Rev.} \textbf{\bibinfo{volume}{D51}},
  \bibinfo{pages}{6177} (\bibinfo{year}{1995}), \eprint{hep-ph/9410280}.

\bibitem[{\citenamefont{Colangelo and De~Fazio}(1998)}]{Colangelo:1997rp}
\bibinfo{author}{\bibfnamefont{P.}~\bibnamefont{Colangelo}} \bibnamefont{and}
  \bibinfo{author}{\bibfnamefont{F.}~\bibnamefont{De~Fazio}},
  \bibinfo{journal}{Eur.Phys.J.} \textbf{\bibinfo{volume}{C4}},
  \bibinfo{pages}{503} (\bibinfo{year}{1998}), \eprint{hep-ph/9706271}.

\bibitem[{\citenamefont{Khodjamirian et~al.}(1999)\citenamefont{Khodjamirian,
  Ruckl, Weinzierl, and Yakovlev}}]{Khodjamirian:1999hb}
\bibinfo{author}{\bibfnamefont{A.}~\bibnamefont{Khodjamirian}},
  \bibinfo{author}{\bibfnamefont{R.}~\bibnamefont{Ruckl}},
  \bibinfo{author}{\bibfnamefont{S.}~\bibnamefont{Weinzierl}},
  \bibnamefont{and} \bibinfo{author}{\bibfnamefont{O.~I.}
  \bibnamefont{Yakovlev}}, \bibinfo{journal}{Phys.Lett.}
  \textbf{\bibinfo{volume}{B457}}, \bibinfo{pages}{245} (\bibinfo{year}{1999}),
  \eprint{hep-ph/9903421}.

\bibitem[{\citenamefont{Colangelo
  et~al.}(1994{\natexlab{b}})\citenamefont{Colangelo, De~Fazio, and
  Nardulli}}]{Colangelo:1994jc}
\bibinfo{author}{\bibfnamefont{P.}~\bibnamefont{Colangelo}},
  \bibinfo{author}{\bibfnamefont{F.}~\bibnamefont{De~Fazio}}, \bibnamefont{and}
  \bibinfo{author}{\bibfnamefont{G.}~\bibnamefont{Nardulli}},
  \bibinfo{journal}{Phys.Lett.} \textbf{\bibinfo{volume}{B334}},
  \bibinfo{pages}{175} (\bibinfo{year}{1994}{\natexlab{b}}),
  \eprint{hep-ph/9406320}.

\bibitem[{\citenamefont{Abada et~al.}(2002)\citenamefont{Abada, Becirevic,
  Boucaud, Herdoiza, Leroy et~al.}}]{Abada:2002xe}
\bibinfo{author}{\bibfnamefont{A.}~\bibnamefont{Abada}},
  \bibinfo{author}{\bibfnamefont{D.}~\bibnamefont{Becirevic}},
  \bibinfo{author}{\bibfnamefont{P.}~\bibnamefont{Boucaud}},
  \bibinfo{author}{\bibfnamefont{G.}~\bibnamefont{Herdoiza}},
  \bibinfo{author}{\bibfnamefont{J.}~\bibnamefont{Leroy}},
  \bibnamefont{et~al.}, \bibinfo{journal}{Phys.Rev.}
  \textbf{\bibinfo{volume}{D66}}, \bibinfo{pages}{074504}
  (\bibinfo{year}{2002}), \eprint{hep-ph/0206237}.

\bibitem[{\citenamefont{Becirevic and Haas}(2011)}]{Becirevic:2009xp}
\bibinfo{author}{\bibfnamefont{D.}~\bibnamefont{Becirevic}} \bibnamefont{and}
  \bibinfo{author}{\bibfnamefont{B.}~\bibnamefont{Haas}},
  \bibinfo{journal}{Eur.Phys.J.} \textbf{\bibinfo{volume}{C71}},
  \bibinfo{pages}{1734} (\bibinfo{year}{2011}), \eprint{0903.2407}.

\bibitem[{\citenamefont{Becirevic and Sanfilippo}(2012)}]{Becirevic:2012pf}
\bibinfo{author}{\bibfnamefont{D.}~\bibnamefont{Becirevic}} \bibnamefont{and}
  \bibinfo{author}{\bibfnamefont{F.}~\bibnamefont{Sanfilippo}}
  (\bibinfo{year}{2012}), \eprint{1210.5410}.

\bibitem[{\citenamefont{Brodsky and Hiller}(1992)}]{Brodsky:1992px}
\bibinfo{author}{\bibfnamefont{S.~J.} \bibnamefont{Brodsky}} \bibnamefont{and}
  \bibinfo{author}{\bibfnamefont{J.~R.} \bibnamefont{Hiller}},
  \bibinfo{journal}{Phys.Rev.} \textbf{\bibinfo{volume}{D46}},
  \bibinfo{pages}{2141} (\bibinfo{year}{1992}).

\bibitem[{\citenamefont{Wilcox et~al.}(1992)\citenamefont{Wilcox, Draper, and
  Liu}}]{Wilcox:1991cq}
\bibinfo{author}{\bibfnamefont{W.}~\bibnamefont{Wilcox}},
  \bibinfo{author}{\bibfnamefont{T.}~\bibnamefont{Draper}}, \bibnamefont{and}
  \bibinfo{author}{\bibfnamefont{K.-F.} \bibnamefont{Liu}},
  \bibinfo{journal}{Phys. Rev.} \textbf{\bibinfo{volume}{D46}},
  \bibinfo{pages}{1109} (\bibinfo{year}{1992}), \eprint{hep-lat/9205015}.

\bibitem[{\citenamefont{Aoki et~al.}(2009)}]{Aoki:2008sm}
\bibinfo{author}{\bibfnamefont{S.}~\bibnamefont{Aoki}} \bibnamefont{et~al.}
  (\bibinfo{collaboration}{PACS-CS}), \bibinfo{journal}{Phys. Rev.}
  \textbf{\bibinfo{volume}{D79}}, \bibinfo{pages}{034503}
  (\bibinfo{year}{2009}), \eprint{0807.1661}.

\bibitem[{\citenamefont{El-Khadra et~al.}(1997)\citenamefont{El-Khadra,
  Kronfeld, and Mackenzie}}]{ElKhadra:1996mp}
\bibinfo{author}{\bibfnamefont{A.~X.} \bibnamefont{El-Khadra}},
  \bibinfo{author}{\bibfnamefont{A.~S.} \bibnamefont{Kronfeld}},
  \bibnamefont{and} \bibinfo{author}{\bibfnamefont{P.~B.}
  \bibnamefont{Mackenzie}}, \bibinfo{journal}{Phys. Rev.}
  \textbf{\bibinfo{volume}{D55}}, \bibinfo{pages}{3933} (\bibinfo{year}{1997}),
  \eprint{hep-lat/9604004}.

\bibitem[{\citenamefont{Bali et~al.}(2011)\citenamefont{Bali, Collins, and
  Ehmann}}]{Bali:2011rd}
\bibinfo{author}{\bibfnamefont{G.~S.} \bibnamefont{Bali}},
  \bibinfo{author}{\bibfnamefont{S.}~\bibnamefont{Collins}}, \bibnamefont{and}
  \bibinfo{author}{\bibfnamefont{C.}~\bibnamefont{Ehmann}},
  \bibinfo{journal}{Phys.Rev.} \textbf{\bibinfo{volume}{D84}},
  \bibinfo{pages}{094506} (\bibinfo{year}{2011}), \eprint{1110.2381}.

\bibitem[{\citenamefont{Burch et~al.}(2010)\citenamefont{Burch, DeTar,
  Di~Pierro, El-Khadra, Freeland et~al.}}]{Burch:2009az}
\bibinfo{author}{\bibfnamefont{T.}~\bibnamefont{Burch}},
  \bibinfo{author}{\bibfnamefont{C.}~\bibnamefont{DeTar}},
  \bibinfo{author}{\bibfnamefont{M.}~\bibnamefont{Di~Pierro}},
  \bibinfo{author}{\bibfnamefont{A.}~\bibnamefont{El-Khadra}},
  \bibinfo{author}{\bibfnamefont{E.}~\bibnamefont{Freeland}},
  \bibnamefont{et~al.}, \bibinfo{journal}{Phys.Rev.}
  \textbf{\bibinfo{volume}{D81}}, \bibinfo{pages}{034508}
  (\bibinfo{year}{2010}), \eprint{0912.2701}.

\bibitem[{\citenamefont{Ali~Khan et~al.}(2002)\citenamefont{Ali~Khan, Aoki,
  Boyd, Burkhalter, Ejiri, Fukugita, Hashimoto, Ishizuka, Iwasaki, Kanaya
  et~al.}}]{AliKhan:2001tx}
\bibinfo{author}{\bibfnamefont{A.}~\bibnamefont{Ali~Khan}},
  \bibinfo{author}{\bibfnamefont{S.}~\bibnamefont{Aoki}},
  \bibinfo{author}{\bibfnamefont{G.}~\bibnamefont{Boyd}},
  \bibinfo{author}{\bibfnamefont{R.}~\bibnamefont{Burkhalter}},
  \bibinfo{author}{\bibfnamefont{S.}~\bibnamefont{Ejiri}},
  \bibinfo{author}{\bibfnamefont{M.}~\bibnamefont{Fukugita}},
  \bibinfo{author}{\bibfnamefont{S.}~\bibnamefont{Hashimoto}},
  \bibinfo{author}{\bibfnamefont{N.}~\bibnamefont{Ishizuka}},
  \bibinfo{author}{\bibfnamefont{Y.}~\bibnamefont{Iwasaki}},
  \bibinfo{author}{\bibfnamefont{K.}~\bibnamefont{Kanaya}},
  \bibnamefont{et~al.}, \bibinfo{journal}{Phys. Rev. D}
  \textbf{\bibinfo{volume}{65}}, \bibinfo{pages}{054505}
  (\bibinfo{year}{2002}).

\bibitem[{\citenamefont{O'Connell et~al.}(1997)\citenamefont{O'Connell, Pearce,
  Thomas, and Williams}}]{OConnell:1995wf}
\bibinfo{author}{\bibfnamefont{H.~B.} \bibnamefont{O'Connell}},
  \bibinfo{author}{\bibfnamefont{B.}~\bibnamefont{Pearce}},
  \bibinfo{author}{\bibfnamefont{A.~W.} \bibnamefont{Thomas}},
  \bibnamefont{and} \bibinfo{author}{\bibfnamefont{A.~G.}
  \bibnamefont{Williams}}, \bibinfo{journal}{Prog.Part.Nucl.Phys.}
  \textbf{\bibinfo{volume}{39}}, \bibinfo{pages}{201} (\bibinfo{year}{1997}),
  \eprint{hep-ph/9501251}.

\bibitem[{\citenamefont{Beringer et~al.}(2012)\citenamefont{Beringer, Arguin,
  Barnett, Copic, Dahl, Groom, Lin, Lys, Murayama, Wohl
  et~al.}}]{PhysRevD.86.010001}
\bibinfo{author}{\bibfnamefont{J.}~\bibnamefont{Beringer}},
  \bibinfo{author}{\bibfnamefont{J.~F.} \bibnamefont{Arguin}},
  \bibinfo{author}{\bibfnamefont{R.~M.} \bibnamefont{Barnett}},
  \bibinfo{author}{\bibfnamefont{K.}~\bibnamefont{Copic}},
  \bibinfo{author}{\bibfnamefont{O.}~\bibnamefont{Dahl}},
  \bibinfo{author}{\bibfnamefont{D.~E.} \bibnamefont{Groom}},
  \bibinfo{author}{\bibfnamefont{C.~J.} \bibnamefont{Lin}},
  \bibinfo{author}{\bibfnamefont{J.}~\bibnamefont{Lys}},
  \bibinfo{author}{\bibfnamefont{H.}~\bibnamefont{Murayama}},
  \bibinfo{author}{\bibfnamefont{C.~G.} \bibnamefont{Wohl}},
  \bibnamefont{et~al.} (\bibinfo{collaboration}{Particle Data Group}),
  \bibinfo{journal}{Phys. Rev. D} \textbf{\bibinfo{volume}{86}},
  \bibinfo{pages}{010001} (\bibinfo{year}{2012}).

\bibitem[{\citenamefont{Bracco et~al.}(2001)\citenamefont{Bracco, Chiapparini,
  Lozea, Navarra, and Nielsen}}]{Bracco:2001dj}
\bibinfo{author}{\bibfnamefont{M.}~\bibnamefont{Bracco}},
  \bibinfo{author}{\bibfnamefont{M.}~\bibnamefont{Chiapparini}},
  \bibinfo{author}{\bibfnamefont{A.}~\bibnamefont{Lozea}},
  \bibinfo{author}{\bibfnamefont{F.}~\bibnamefont{Navarra}}, \bibnamefont{and}
  \bibinfo{author}{\bibfnamefont{M.}~\bibnamefont{Nielsen}},
  \bibinfo{journal}{Phys.Lett.} \textbf{\bibinfo{volume}{B521}},
  \bibinfo{pages}{1} (\bibinfo{year}{2001}), \eprint{hep-ph/0108223}.

\bibitem[{\citenamefont{Bracco et~al.}(2008)\citenamefont{Bracco, Chiapparini,
  Navarra, and Nielsen}}]{Bracco:2007sg}
\bibinfo{author}{\bibfnamefont{M.}~\bibnamefont{Bracco}},
  \bibinfo{author}{\bibfnamefont{M.}~\bibnamefont{Chiapparini}},
  \bibinfo{author}{\bibfnamefont{F.}~\bibnamefont{Navarra}}, \bibnamefont{and}
  \bibinfo{author}{\bibfnamefont{M.}~\bibnamefont{Nielsen}},
  \bibinfo{journal}{Phys.Lett.} \textbf{\bibinfo{volume}{B659}},
  \bibinfo{pages}{559} (\bibinfo{year}{2008}), \eprint{0710.1878}.

\bibitem[{\citenamefont{Bracco et~al.}(2012)\citenamefont{Bracco, Chiapparini,
  Navarra, and Nielsen}}]{Bracco:2011pg}
\bibinfo{author}{\bibfnamefont{M.}~\bibnamefont{Bracco}},
  \bibinfo{author}{\bibfnamefont{M.}~\bibnamefont{Chiapparini}},
  \bibinfo{author}{\bibfnamefont{F.}~\bibnamefont{Navarra}}, \bibnamefont{and}
  \bibinfo{author}{\bibfnamefont{M.}~\bibnamefont{Nielsen}},
  \bibinfo{journal}{Prog.Part.Nucl.Phys.} \textbf{\bibinfo{volume}{67}},
  \bibinfo{pages}{1019} (\bibinfo{year}{2012}), \eprint{1104.2864}.

\bibitem[{\citenamefont{El-Bennich et~al.}(2012)\citenamefont{El-Bennich,
  Krein, Chang, Roberts, and Wilson}}]{ElBennich:2011py}
\bibinfo{author}{\bibfnamefont{B.}~\bibnamefont{El-Bennich}},
  \bibinfo{author}{\bibfnamefont{G.}~\bibnamefont{Krein}},
  \bibinfo{author}{\bibfnamefont{L.}~\bibnamefont{Chang}},
  \bibinfo{author}{\bibfnamefont{C.~D.} \bibnamefont{Roberts}},
  \bibnamefont{and} \bibinfo{author}{\bibfnamefont{D.~J.}
  \bibnamefont{Wilson}}, \bibinfo{journal}{Phys.Rev.}
  \textbf{\bibinfo{volume}{D85}}, \bibinfo{pages}{031502}
  (\bibinfo{year}{2012}), \eprint{1111.3647}.

\bibitem[{\citenamefont{Casalbuoni et~al.}(1992)\citenamefont{Casalbuoni,
  Deandrea, Di~Bartolomeo, Gatto, Feruglio et~al.}}]{Casalbuoni:1992gi}
\bibinfo{author}{\bibfnamefont{R.}~\bibnamefont{Casalbuoni}},
  \bibinfo{author}{\bibfnamefont{A.}~\bibnamefont{Deandrea}},
  \bibinfo{author}{\bibfnamefont{N.}~\bibnamefont{Di~Bartolomeo}},
  \bibinfo{author}{\bibfnamefont{R.}~\bibnamefont{Gatto}},
  \bibinfo{author}{\bibfnamefont{F.}~\bibnamefont{Feruglio}},
  \bibnamefont{et~al.}, \bibinfo{journal}{Phys.Lett.}
  \textbf{\bibinfo{volume}{B292}}, \bibinfo{pages}{371} (\bibinfo{year}{1992}),
  \eprint{hep-ph/9209248}.

\bibitem[{\citenamefont{Woloshyn}(1986)}]{Woloshyn:1985vd}
\bibinfo{author}{\bibfnamefont{R.}~\bibnamefont{Woloshyn}},
  \bibinfo{journal}{Phys.Rev.} \textbf{\bibinfo{volume}{D34}},
  \bibinfo{pages}{605} (\bibinfo{year}{1986}).

\bibitem[{\citenamefont{El-Bennich et~al.}(2008)\citenamefont{El-Bennich,
  de~Melo, Loiseau, Dedonder, and Frederico}}]{ElBennich:2008qa}
\bibinfo{author}{\bibfnamefont{B.}~\bibnamefont{El-Bennich}},
  \bibinfo{author}{\bibfnamefont{J.}~\bibnamefont{de~Melo}},
  \bibinfo{author}{\bibfnamefont{B.}~\bibnamefont{Loiseau}},
  \bibinfo{author}{\bibfnamefont{J.-P.} \bibnamefont{Dedonder}},
  \bibnamefont{and}
  \bibinfo{author}{\bibfnamefont{T.}~\bibnamefont{Frederico}},
  \bibinfo{journal}{Braz.J.Phys.} \textbf{\bibinfo{volume}{38}},
  \bibinfo{pages}{465} (\bibinfo{year}{2008}), \eprint{0805.0768}.

\bibitem[{\citenamefont{Hwang}(2010)}]{Hwang:2009qz}
\bibinfo{author}{\bibfnamefont{C.-W.} \bibnamefont{Hwang}},
  \bibinfo{journal}{Phys.Rev.} \textbf{\bibinfo{volume}{D81}},
  \bibinfo{pages}{054022} (\bibinfo{year}{2010}), \eprint{0910.0145}.

\bibitem[{\citenamefont{Alexandrou et~al.}(2012)\citenamefont{Alexandrou,
  Hadjiyiannakou, Koutsou, O'Cais, and Strelchenko}}]{Alexandrou:2012zz}
\bibinfo{author}{\bibfnamefont{C.}~\bibnamefont{Alexandrou}},
  \bibinfo{author}{\bibfnamefont{K.}~\bibnamefont{Hadjiyiannakou}},
  \bibinfo{author}{\bibfnamefont{G.}~\bibnamefont{Koutsou}},
  \bibinfo{author}{\bibfnamefont{A.}~\bibnamefont{O'Cais}}, \bibnamefont{and}
  \bibinfo{author}{\bibfnamefont{A.}~\bibnamefont{Strelchenko}},
  \bibinfo{journal}{Comput.Phys.Commun.} \textbf{\bibinfo{volume}{183}},
  \bibinfo{pages}{1215} (\bibinfo{year}{2012}), \eprint{1108.2473}.

\bibitem[{\citenamefont{Edwards and Joo}(2005)}]{Edwards:2004sx}
\bibinfo{author}{\bibfnamefont{R.~G.} \bibnamefont{Edwards}} \bibnamefont{and}
  \bibinfo{author}{\bibfnamefont{B.}~\bibnamefont{Joo}}
  (\bibinfo{collaboration}{SciDAC Collaboration, LHPC Collaboration, UKQCD
  Collaboration}), \bibinfo{journal}{Nucl.Phys.Proc.Suppl.}
  \textbf{\bibinfo{volume}{140}}, \bibinfo{pages}{832} (\bibinfo{year}{2005}),
  \eprint{hep-lat/0409003}.

\end{thebibliography}
\end{document}